\documentclass[lineno*]{jfm}

\usepackage{graphicx}
\usepackage{newtxtext}
\usepackage{newtxmath}
\usepackage{natbib}
\usepackage{hyperref}
\hypersetup{colorlinks,allcolors=blue}

\newcommand{\RomanNumeralCaps}[1]
\title{Nonlinear bubble behaviours of compressible Rayleigh--Taylor instability with isothermal stratification in cylindrical geometry}

\author{Ming Yuan\aff{1},
    Zhiye Zhao\aff{1,2}	\corresp{\email{zzy12@ustc.edu.cn}},
    Luoqin Liu\aff{1},
    Pei Wang\aff{3},
    Nan-Sheng Liu\aff{1}
    \and Xi-Yun Lu\aff{1}}

\affiliation{\aff{1}Department of Modern Mechanics, University of Science and Technology of China, Hefei, Anhui 230026, PR China
\aff{2}Department of Mechanical and Aerospace Engineering, The Hong Kong University of Science and Technology, Hong Kong, China
\aff{3}Institute of Applied Physics and Computational Mathematics, Beijing 100094, PR China}

\begin{document}
\maketitle

\begin{abstract}
	Nonlinear evolutions of two-dimensional single-mode compressible Rayleigh--Taylor instability (RTI) with isothermal stratification are investigated in cylindrical geometry via direct numerical simulation for different Atwood numbers ($A_T=0.1-0.9$) and Mach numbers ($Ma=0.1-0.9$). It is found that the nonlinear bubble growth involves the effects of density stratification, vorticity accumulation and flow compressibility and shows considerable differences between convergent (acceleration acting radially inward) and divergent (acceleration acting radially outward) cases. Specifically, the density stratification leads to non-acceleration at low $A_T$ and high $Ma$. The accelerations in convergent cases are dominated by vorticity accumulation at low $A_T$ and low $Ma$ and by flow compressibility at high $A_T$ and high $Ma$ whereas the accelerations in divergent cases are purely induced by flow compressibility at high $A_T$ and high $Ma$. Based on the nonlinear theory of incompressible cylindrical RTI with uniform-density background~(Zhao et al., \emph{J. Fluid Mech.}, vol. 900, 2020, A24), an improved model is proposed by taking the density variation, vorticity accumulation and flow compressibility into consideration. This model is verified by numerical results and well reproduces the bubble evolution for different $A_T$ and $Ma$ from linear to highly nonlinear regimes.
\end{abstract}


\section{Introduction}\label{sec:intro}
    Rayleigh--Taylor (RT) instability arises at a perturbed interface between two fluids with different densities when the heavy fluid is accelerated by the light fluid in the presence of an external acceleration~\citep{Rayleigh1883,Taylor1950}. With the development of RT instability (RTI), the light fluid rising into the heavy one and the heavy fluid sinking into the light one result in the formation of bubble-like and spike-like finger structures~\citep{Zhou20171,Zhou20172}, respectively. These structures are of both fundamental interest and practical importance in many phenomena and applications, including geological flows~\citep{Houseman1997}, oceanic flows~\citep{Livescu2013}, astrophysical flows~\citep{Bell2004,Isobe2005}, premixed combustion~\citep{Chertkov2009,Hicks2014,Sykes2021}, nuclear fusion~\citep{Lindl2014} and explosive detonation~\citep{Balakrishnan2010}. Many of these systems often intimately involve background stratification due to the presence of an acceleration field~\citep{Livescu2013}, which results in particularly significant flow compressibility induced by the variation of the fluid density. Thus, stratified compressible RTI development has attracted widespread attention through theoretical and numerical studies over the past decades.

	Great advancement has emerged on the stratified RTI development in planar geometry from linear to nonlinear saturation and highly nonlinear regimes. Specifically, \citet{Livescu2004} performed a linear stability analysis of compressible RTI in the linear regime and concluded that the density stratification has a stabilizing effect on early instability growth by reducing the average local Atwood number ($A_T$), a parameter characterising the density ratio. In the nonlinear saturation regime~\citep{Reckinger2016,Luo2020,Fu2022}, density stratification at small (large) $A_T$ makes bubble velocity lower (higher) than the saturation value obtained from the corresponding incompressible uniform-density counterpart~\citep{Layzer1955,Goncharov2002}. This is attributed to the fact that the effect of density stratification plays a dominantly stabilizing role at small $A_T$ while flow compressibility becomes dominating and acts as a destabilizing role at high $A_T$ due to the compression of heavy fluid exerted by the rising bubble~\citep{Luo2020,Fu2022}. As for the highly nonlinear bubble behaviours after nonlinear saturation regime, the bubble is re-accelerated to a velocity well above the saturation value due to vorticity accumulation inside the bubble in incompressible and weakly compressible RTI with sufficiently high Reynolds numbers~\citep{Betti2006,Bian2020}. And the Betti--Sanz model~\citep{Betti2006} by introducing a vorticity term to the saturation velocity model can successfully describe the bubble re-acceleration (RA) behaviours. Recently, \citet{Fu2023} examined the bubble RA behaviours of stratified RTI with various $A_T$ and Mach numbers ($Ma$), a parameter characterising the stratification strength, and found that flow compressibility would dominate the bubble RA behaviours at high $A_T$ and high $Ma$. An improved model was proposed by introducing an additional term characterising the flow compressibility to the Betti--Sanz model~\citep{Betti2006} and well captured the bubble RA dynamics of stratified RTI. Furthermore, the density stratification weakens the conversion from potential energy to kinetic energy while flow compressibility is critical to convert the internal energy into the kinetic energy in compressible RT turbulence~\citep{Zhao2020a,Luo2021,Zhao2022,Luo2022}.

	Compared with planar RTI, the RTI in a convergent geometry is of more particular relevance to supernova explosions~\citep{Hester2008} and inertial confinement fusion (ICF)~\citep{Kishony2001,Betti2016}, and thus is of more practical interest. Cylindrical geometry which involves principal effects of convergent geometries has been widely used as a natural choice to study the convergent effects on hydrodynamic instability evolution~\citep{Bell1951,Sakagami1990,Weir1998,Wang2013,Zhao2020b,Zhao2021,Wu2021,Ge2022,Yuan2023}. In the incompressible limit, cylindrical RTI increases exponentially with a growth rate $\Gamma=\sqrt{A_T |g|n/r_0}$ for both convergent (acceleration acting radially inward $g<0$) and divergent (acceleration acting radially outward $g>0$) cases in the linear regime~\citep{Wang2013}, where $n$ is the number of cosinoidal perturbation waves and $r_0$ is the radius of the unperturbed interface. To further elucidate the bubble evolution of convergent and divergent cases, \citet{Zhao2020b} employed a nonlinear theory based on potential flow assumption and presented an analytical model covering bubble evolution at arbitrary $A_T$ from linear to nonlinear regimes. It is revealed from this nonlinear theory that the bubble velocity in a convergent (divergent) case undergoes a uniform acceleration (deceleration) in the nonlinear regime instead of keeping an asymptotic value like planar cases. In such instances, both theoretical analysis and simulation~\citep{Wang2013,Zhao2020b} have been considered near the incompressible limit with negligible background stratification while stratified RTI has received much less attention in cylindrical geometry. It is found from the work of \citet{Yu2008} that density stratification also plays a stabilizing role in cylindrical RTI growth in the linear regime, similar to the results of planar cases~\citep{Livescu2004}. However, nonlinear and highly nonlinear bubble behaviors of stratified cylindrical RTI have not been systematically investigated yet.
	
	Motivated by the aforementioned findings, nonlinear and highly nonlinear evolutions of stratified compressible RTI in cylindrical geometry are worthy of further investigation. This paper focuses on the cylindrical RTI growth characteristics of convergent and divergent cases, with special interest directed to answer the following questions. What will the bubble dynamics of compressible cylindrical RTI like and how will flow compressibility and vorticity accumulation set in the highly nonlinear regime? Towards this goal, direct numerical simulation (DNS) of two-dimensional single-mode stratified compressible RTI is performed in cylindrical geometry over a range of $A_T$ and $Ma$. The remainder of this paper is organized as follows. The numerical strategy used to simulate the instability evolution is briefly described in \S~2. The general features of bubble behaviors are discussed and an improved model is proposed to characterise the bubble growth in \S~3. Finally, conclusions and recommendations for future work are addressed in \S~4.

\section{Numerical simulations}\label{sec:simu}
    \subsection{Governing equations}
		Direct numerical simulation has been performed on stratified compressible RTI in cylindrical geometry to study the bubble behaviours. According to previous studies~\citep{Zhao2020b,Zhao2021}, the radius of the unperturbed interface $r^*_0$, the pressure $p^*_I$ and density $\rho^*_I$ at the initial interface are chosen as the characteristic scales. Here, the characteristic velocity and temperature are described, respectively, as $u^*_I=\sqrt{p^*_I/\rho^*_I}$ and $T^*_I=p^*_I M^*_I/(R^* \rho^*_I)$, where $R^*$ is the universal gas constant and the molar mass is $M^*_I=(M^*_h+M^*_l)/2$ with $M^*_h$ and $M^*_l$ representing the molar masses of heavy and light fluids, respectively. Hereafter, the superscript `$*$' denotes dimensional physical quantities and the subscript `$I$' corresponds to the quantities at the initial interface. Thus, the non-dimensionalized governing equations in cylindrical coordinates $(r,\phi)$ are
		
		\begin{equation}
			\displaystyle\frac{\p \rho}{\p t}+\bnabla\bcdot(\rho \boldsymbol{u})=0,
			\label{eq:Gov1}
		\end{equation}
		
		\begin{equation}
			\displaystyle\frac{\p (\rho \boldsymbol{u})}{\p t}+\bnabla\bcdot(\rho \boldsymbol{uu})=
			-\bnabla p+\frac{1}{Re}\bnabla\bcdot\boldsymbol{\tau}+\frac{\rho}{Fr}\boldsymbol{e}_r,
			\label{eq:Gov2}
		\end{equation}
		
		\begin{equation}
			\displaystyle\frac{\p (\rho E)}{\p t}+\bnabla\bcdot[(\rho E+p)\boldsymbol{u}]=
			\frac{1}{Re}\bnabla\bcdot(\boldsymbol{\tau\bcdot u})-\frac{1}{RePr}\bnabla\bcdot\boldsymbol{q}_c
			-\frac{1}{ReSc}\bnabla\bcdot\boldsymbol{q}_d+\frac{\rho}{Fr}\boldsymbol{u}\bcdot\boldsymbol{e}_r,
			\label{eq:Gov3}
		\end{equation}
		
		\begin{equation}
			\displaystyle\frac{\p (\rho Y_h)}{\p t}+\bnabla\bcdot(\rho Y_h \boldsymbol{u})=
			-\frac{1}{Re Sc}\bnabla\bcdot\boldsymbol J_h,
			\label{eq:Gov4}
		\end{equation}
		where $\rho$ is the fluid density; $\boldsymbol{u}=(u_r,u_\phi)$ denotes the velocity vector; $p$ is the pressure; $\boldsymbol{e}_r$ is the unit vector in the radial direction; $E=C_vT+\boldsymbol{u}\bcdot\boldsymbol{u}/2$ denotes the specific total energy with $C_v$ being the specific heat at constant volume and $T$ the temperature; $Y_h=\rho_h/\rho$ is the species mass fraction of heavy fluid and $Y_l=1-Y_h$ is the species mass fraction of light fluid; and the symbol $\bnabla$ denotes the vector-differentiation operator. The stress tensor is obtained as $\boldsymbol\tau=2\mu\boldsymbol S-2\mu/3(\bnabla\bcdot\boldsymbol u)\boldsymbol\delta$, where the dynamic viscosity $\mu=T^{3/2}(1+c)/(T+c)$ is computed by the Sutherland law with $c=T_s^*/T_r^*$ where the constant temperature $T_s^*=124$~$\rm{K}$ and the reference temperature $T_r^*=273.15$~$\rm{K}$, $\boldsymbol S=(\bnabla\boldsymbol{u}+(\bnabla\boldsymbol{u}) ^{T})/2$ is the strain-rate tensor and $\boldsymbol\delta$ represents the unit tensor. The heat fluxes due to heat conduction ($\boldsymbol{q}_c$) and interspecies enthalpy diffusion ($\boldsymbol{q}_d$) are given by $\boldsymbol{q}_c=-\gamma/\left[ M_I(\gamma-1)\right] \kappa\bnabla T$ and $\boldsymbol{q}_d=\sum h_i\boldsymbol J_i$~$ (i=h, l)$, respectively, where $\gamma$ is the ratio of specific heats, $M_I$ is the molar mass at the initial interface, $\kappa$ is the heat conduction coefficient, $h_i$ is the enthalpy, $\boldsymbol J_i=-\rho D \bnabla Y_i$ is the diffusive mass flux obtained by the Fick law and $D$ is the diffusion coefficient. The above governing equations are closed with the non-dimensionalized ideal gas equation of state, i.e. $p=\rho T /M$, where $M$ is the molar mass.
		
		Following recent work~\citep{Ge2022}, the density and pressure of the mixture are obtained by the summation of each species, while the temperature is equal for each species of the mixture. Therefore, the molecular mass of the mixture is given by $M=(\sum Y_i/M_i)^{-1}$, where $M_i$ is the molecular mass of the $i$th species. The quantities describing the physical properties of the mixture, such as the dynamic viscosity $\mu$, the diffusion coefficient $D$, the heat conduction coefficient $\kappa$, the specific heat at constant pressure $C_{p}$ and the specific heat at constant volume $C_{v}$, are calculated by the linear combinations of each species weighted with their mass fractions~\citep{Reckinger2016}.
		
		The non-dimensional parameters in~(\ref{eq:Gov1})--(\ref{eq:Gov4}) are the Reynolds, Prandtl, Schmidt and Froude numbers defined, respectively, as
		\refstepcounter{equation}
		$$			
			Re=\frac{\rho_I^* u_I^* r^*_0}{\mu_I^*},
			\quad Pr=\frac{C_{p,I}^* \mu_I^*}{\kappa_I^*},
			\quad Sc=\frac{\mu_I^*}{\rho_I^* D_I^*},
			\quad Fr=\frac{p^*_I/\rho^*_I}{r_0^*g^*},
			\eqno{(\theequation{\mathit{a}-\mathit{d}})}\label{eq:Nondim}
		$$
		where $g^*$ is the radially external acceleration with $g^*<0$ for convergent cases and $g^*>0$ for divergent cases.
		
	\subsection{Problem set-up and numerical method}
	    In the present study, the shape function of a cosinoidally perturbed interface takes form as $\zeta(\phi)=r_0+\eta_0\cos(n\phi)$. Here, $\eta_0$ represents the initial amplitude and $n$ denotes the number of perturbation waves. Then, the Mach number based on the initial perturbation wavelength $\lambda^*_0=2\pi r_0^*/n$ is defined as $Ma=\sqrt{|g^*|\lambda^*_0/(p^*_I/\rho^*_I)}$ to illustrate the strength of background stratification, following previous studies~\citep{Reckinger2016,Fu2023}. Initially, this compressible RT system keeps hydrostatic equilibrium ($\boldsymbol{u}=0$). By integrating the momentum equation (\ref{eq:Gov2}) with ideal gas equation of state and isothermal assumption ($T=1$)~\citep{Livescu2004,Reckinger2016,Fu2023}, the initial density and pressure fields are yielded as
		\begin{subeqnarray}
			\rho_{h,l}=(1 &\pm& A_T)\exp[\mathrm{sgn}(g^*) Ma^2 (1\pm A_T) \frac{r-\zeta(\phi)}{2\pi/n}],\\[3pt]
			p_{h,l}&=&\exp[\mathrm{sgn}(g^*) Ma^2 (1 \pm A_T) \frac{r-\zeta(\phi)}{2\pi/n}],
			\label{eq:Initial}
		\end{subeqnarray}
		where $A_T=(M^*_h-M^*_l)/(M^*_h+M^*_l)$ is the Atwood number and $\mathrm{sgn} (g^*)$ is the sign function, namely if $g^*>0$, $\mathrm{sgn}(g^*)=1$, and if $g^*<0$, $\mathrm{sgn}(g^*)=-1$. To smooth the interface density jump, the error function is introduced~\citep{Reckinger2016,Bian2020,Fu2023}.

		A wide range of Mach numbers ($Ma=0.1-0.9$) and Atwood numbers ($A_T=0.1-0.9$) is considered to examine the RTI development with various stratification strengths and density ratios. The number of perturbation waves is chosen as $n=8$ according to~\citet{Zhao2020b} and the intial amplitude is set as $\eta_0=0.02\lambda_0$ to satisfy the small-perturbation assumption~\citep{Fu2023}. To make the bubble acceleration behavior in highly nonlinear regime clearly evident as pointed by~\citet{Bian2020}, a sufficiently high perturbation Reynolds number ($Re_p=\rho_I^*\lambda^*_0\sqrt{A_T/(1+A_T)|g^*|\lambda^*_0}/\mu_I^*$) is chosen as $10000$.
		Notably, this Reynolds number is defined by the initial perturbation wavelength $\lambda^*_0$ and a characteristic velocity $V_t=\sqrt{A_T/(1+A_T)|g^*|\lambda^*_0}$ proportional to the saturation velocity obtained from potential flow theory~\citep{Goncharov2002}.
		Other parameters in governing equations~(\ref{eq:Gov1})--(\ref{eq:Gov4}) are fixed as $Pr=0.72$, $Sc=1$ and $\gamma=1.4$. The bubble position and velocity are determined as the point of the bubble tip with the mole fraction of heavy fluid $X_h=[(1-A_T)Y_h]/(1+A_T-2A_TY_h)=0.5$~\citep{Luo2021}. 
		To compare nonlinear bubble evolution under different parameters, the bubble velocity and time in the following discussion are rescaled to ensure that the bubbles have the same scaled velocity when entering the nonlinear regime at the same scaled time~\citep{Bian2020,Fu2023}. The exponential growth rate of cylindrical RTI $\Gamma=\sqrt{A_T |g^*| n/r_0^*}$ is chosen to characterise the time following recent work~\citep{Zhao2020b} and the characteristic velocity $V_t$ in the definition of $Re_p$ is used neutrally to rescale the bubble velocity, referring to the work of~\citet{Bian2020} and~\citet{Fu2023}.

		In addition, all simulations are performed within a two-dimensional circular domain $D = \left\{(r,\phi) | 0.05\leqslant r\leqslant 2.5, 0\leqslant\phi < 2\pi\right\}$. The computational domain has been verified to have a negligible effect on cylindrical RTI evolution in previous simulations~\citep{Zhao2020b,Zhao2021}. It is noted that a micro-hole with a radius of $0.05$ is dug out to avoid a pole singularity at the centre of cylindrical coordinates, following previous treatments~\citep{Zhao2020b,Wu2021,Yuan2023}. The  free-slip (stress-free) conditions for velocities, a zero heat flux condition for temperature and a zero mass flux condition for mass fraction are applied to the interior and exterior boundaries~\citep{Gauthier2017,Bian2020}.

		\begin{figure}
			\centering
			\includegraphics[scale=0.64]{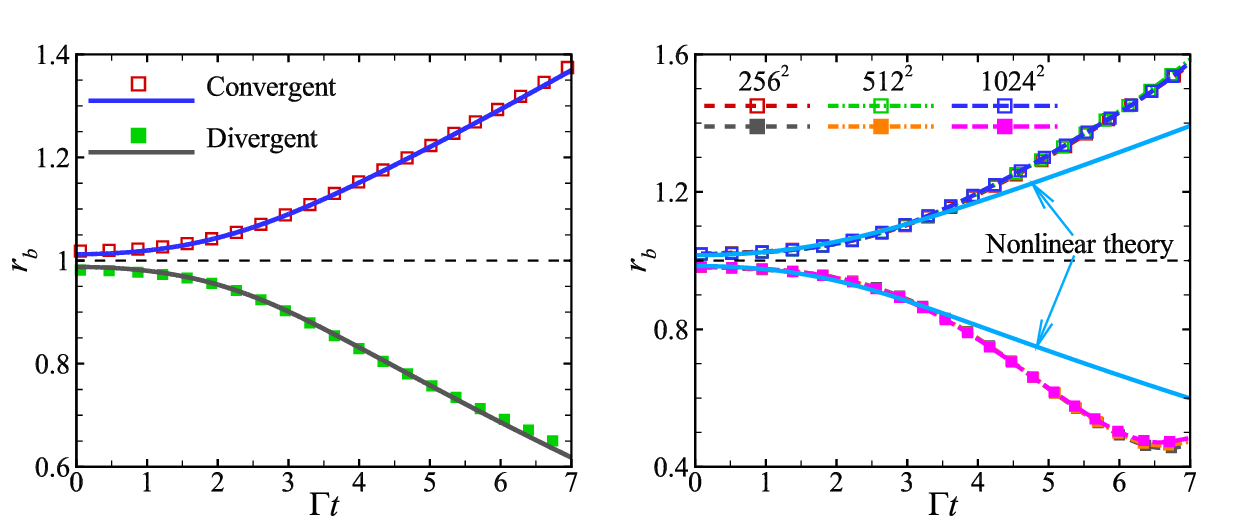}
			\put(-385,140){($a$)}
			\put(-195,140){($b$)}
			\caption{Time-varying positions of bubble tips $r_b$ for convergent ($r_b>1$) and divergent ($r_b<1$) cases at ($a$) $A_T=0.9$ and $Ma=0.1$ with symbols representing the simulation data and at ($b$) $A_T=0.9$ and $Ma=0.9$ with three grid resolutions: $256^2$ (dashed lines), $512^2$ (dot-dashed lines) and $1024^2$ (long-dashed lines). The solid lines denote the nonlinear theory~\citep{Zhao2020b}.}
			\label{fig:Grid}
		\end{figure}

		A numerical algorithm of high-order finite difference schemes is used to solve the governing equations~(\ref{eq:Gov1})--(\ref{eq:Gov4}) in cylindrical coordinates~\citep{Zhao2020b,Zhao2021,Yuan2023}. Specifically, the fifth-order weighted essentially non-oscillatory scheme is implemented to discretize the convective terms. The sixth-order central difference scheme is performed to discretize the viscous terms. The time derivative is approximated by the classical third-order Runge--Kutta method. At a sufficiently weak stratification strength ($Ma=0.1$), the flow can be considered to be nearly the incompressible limit~\citep{Luo2021} and the corresponding bubble growth consequently satisfies the nonlinear theory~\citep{Zhao2020b}. As shown in figure~\ref{fig:Grid}($a$), the time-varying positions of bubble tips of the simulation data at $A_T=0.9$ and $Ma=0.1$ show good consistency with the nonlinear theory~\citep{Zhao2020b} from linear to nonlinear regimes, indicating the reliability of the present numerical settings. Furthermore, grid convergence should be examined in order to accurately capture the bubble growth. Since the flow compressibility are the strongest at $A_T=0.9$ and $Ma=0.9$, we here show the numerical results with different grid resolutions at $A_T=0.9$ and $Ma=0.9$, the most demanding in grid resolution. Figure~\ref{fig:Grid}($b$) shows that the data at three grid resolutions almost collapse together and thus the present simulations are reliable for capturing the essential flow dynamics in cylindrical RTI evolution. To obtain fine flow field characteristics, the following discussion and analysis are obtained on the finest grid ($1024^2$).

\section{Results and discussions}\label{sec:result}
	\subsection{Bubble nonlinear behaviors}

\begin{figure}
			\centering
			\includegraphics[scale=0.64]{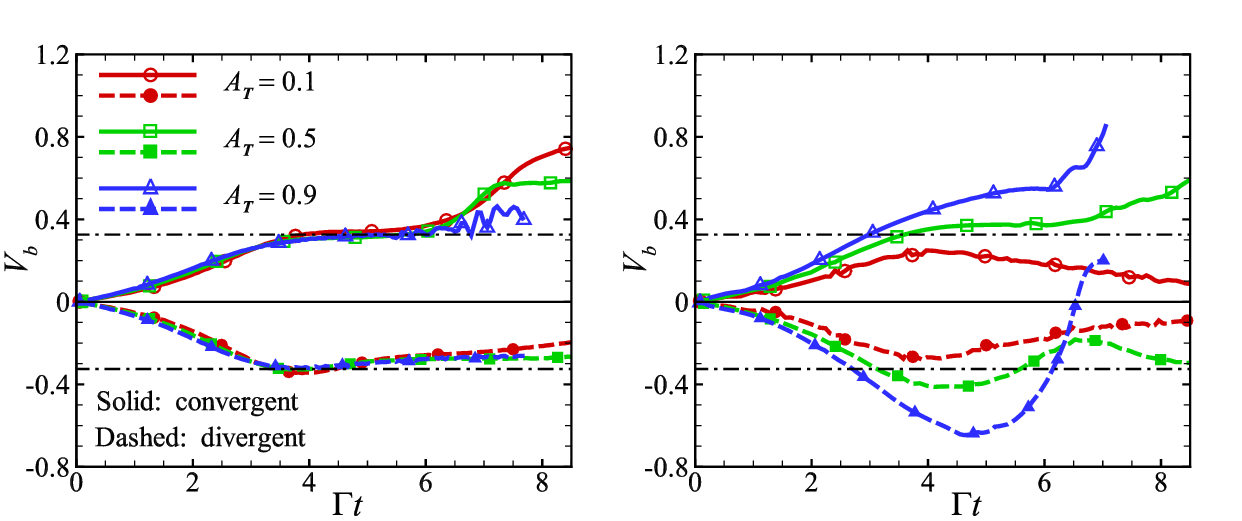}
			\put(-385,140){($a$)}
			\put(-195,140){($b$)}
			\caption{Temporal bubble velocity $V_b$ at different $A_T$ of ($a$) $Ma=0.1$ and of ($b$) $Ma=0.9$. The solid and long-dashed lines in each panel represent the bubble velocities of convergent and divergent cases for $n=8$, respectively. The horizontal black dot-dashed lines in each panel denote the saturation velocity obtained from potential flow theory~\citep{Goncharov2002}.}
			\label{fig:Results}
		\end{figure}

		To investigate the influence of $Ma$ and $A_T$, the time-varying bubble velocity $V_b$ at different $A_T$ for weak ($Ma=0.1$) and strong ($Ma=0.9$) stratification strengths is shown in figure~\ref{fig:Results}.	Similar to the results of recent works~\citep{Bian2020,Fu2023}, the simulations with a higher Atwood number (e.g. $A_T=0.9$) end sooner due to the spike approaching the interior ($g^*<0$) or exterior ($g^*>0$) wall in less time. Bubble velocity $V_b$ increases exponentially for all cases during the linear stage about $\Gamma t<2$ and then shows distinct growth trends for different parameters settings when evolving into the nonlinear regimes. More details and explanations will be given in the following.

		\begin{figure}
			\centering
			\includegraphics[scale=0.53]{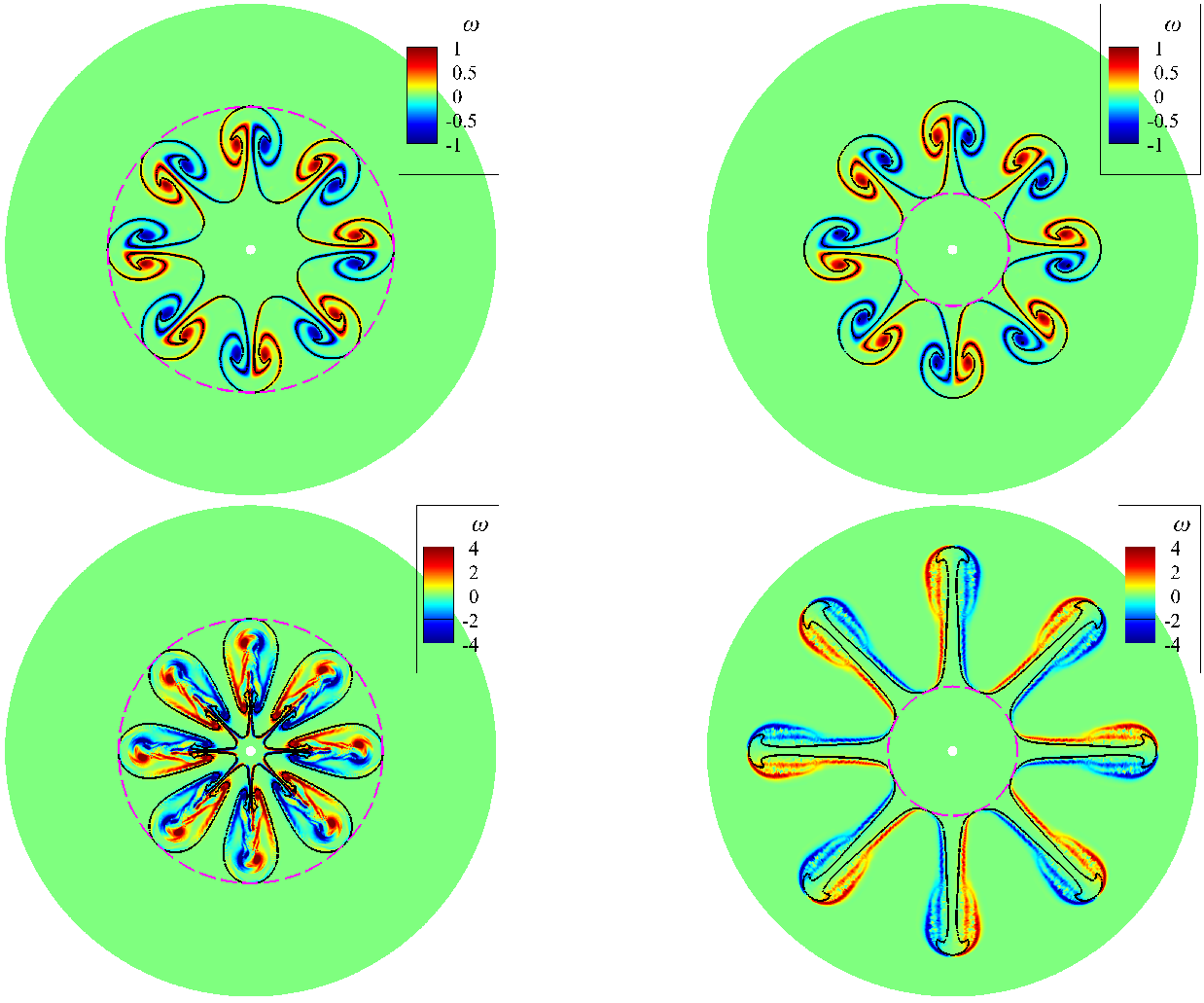}
			\put(-320,240){($a$)}
			\put(-135,240){($b$)}
			\put(-320,105){($c$)}
			\put(-135,105){($d$)}
			\quad
			\centering
			\includegraphics[scale=0.53]{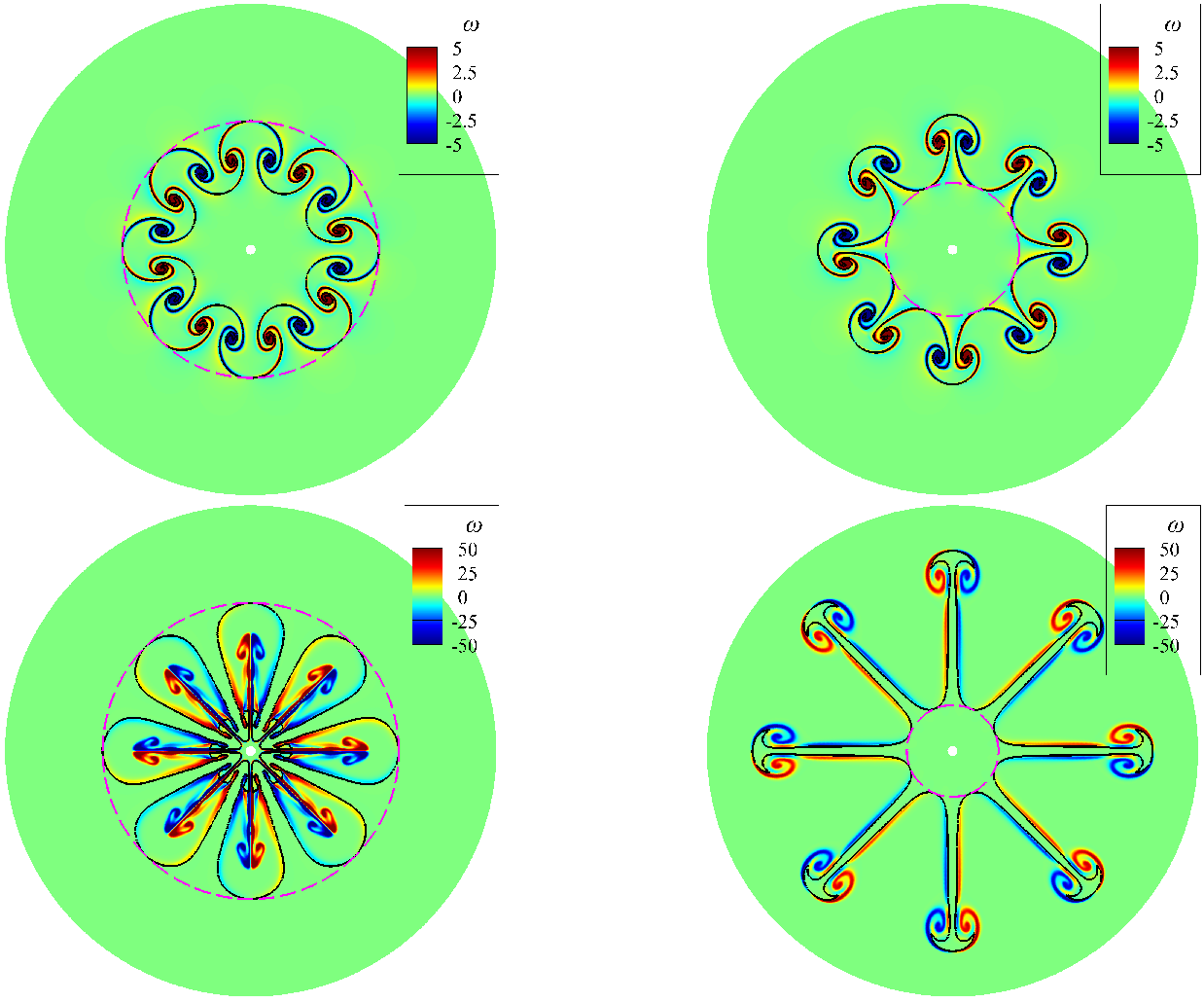}
			\put(-320,240){($e$)}
			\put(-135,240){($f$)}
			\put(-320,105){($g$)}
			\put(-135,105){($h$)}			
			\caption{Vorticity contours for the cases into highly nonlinear stage ($\Gamma t=6.5$) at different $A_T$ with $n=8$ for ($a-d$) $Ma=0.1$ and for $(e-h)$ $Ma=0.9$. The left column ($a,c,e,g$) and right column ($b,d,f,h$) denote convergent and divergent cases, respectively. The first and third rows ($a,b,e,f$) represent cases at $A_T=0.1$ and the second and last rows ($c,d,g,h$) denote cases at $A_T=0.9$. The black solid lines denote the interfaces at $X_h=0.5$. The colored dashed circular lines represent radial positions of bubble tips.}
			\label{fig:Wz}
		\end{figure}

		At $Ma=0.1$, $V_b$ for different $A_T$ reaches the saturation value obtained from potential flow theory~\citep{Goncharov2002} at $\Gamma t\approx4$. After that, the bubble behaviors of convergent cases differ significantly from those of corresponding divergent cases. As for convergent cases, $V_b$ first experiences a slightly uniform acceleration about $\Gamma t<6$ and then undergoes an obvious acceleration where $V_b$ increases rapidly. Previous nonlinear theory~\citep{Zhao2020b} correctly predicts the bubble behavior in the early nonlinear stage ($\Gamma t<6$) but fails in this highly nonlinear phase. Figure~\ref{fig:Wz} illustrates vorticity contours ($\omega=\bnabla\times\boldsymbol{u}$) for various cases into highly nonlinear stage and it can be clearly seen from figure~\ref{fig:Wz}($a,b$) that the vorticity dominates the flow field inside the bubbles for convergent cases at $Ma=0.1$. At this point, both density stratification and flow compressibility are negligibly weak and thus vorticity accumulation inside the bubble is responsible for this acceleration behavior. The main destabilizing mechanism induced by vorticity accumulation inside the bubble is that the vortex pairs inside the bubble approach towards bubble tip, inducing an equivalently centrifugal force to aid bubble growth~\citep{Betti2006}. Furthermore, increasing $A_T$ from $0.1$ to $0.9$ reduces the maximum value of $V_b$, which is consistent with the results of planar RTI~\citep{Bian2020,Luo2021,Fu2023}. Increasing $A_T$ makes sites of vortex generation around the sinking spike drift further away from the bubble tip, so that vortices at higher $A_T$ travel for a longer distance and obtain an attenuated dissipation before entering the bubble tip region for a fixed $Re_p$~\citep{Bian2020}.
		Differently, $V_b$ for divergent cases undergoes a long-time uniform deceleration after reaching the saturation value, as predicted by the nonlinear theory~\citep{Zhao2020b}. Figure~\ref{fig:Wz}($c,d$) shows that the vortex pairs generated from outward spikes are highly far away from inward bubble tips and thus cannot make bubble accelerate. Therefore, the acceleration induced by vorticity accumulation cannot be observed for divergent cases. These results indicate that the acceleration mechanism of vorticity accumulation is fundamentally different in convergent and divergent cases.

\begin{figure}
			\centering
			\includegraphics[scale=0.64]{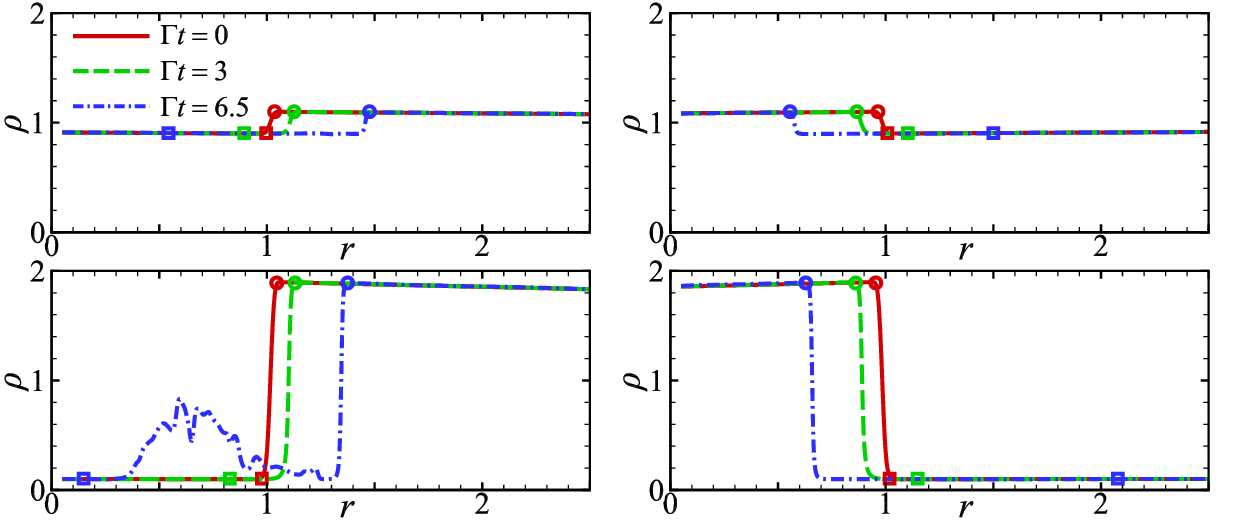}
			\put(-385,154){($a$)}
			\put(-195,154){($b$)}
			\put(-385,74){($c$)}
			\put(-195,74){($d$)}
			\quad
			\centering
			\includegraphics[scale=0.64]{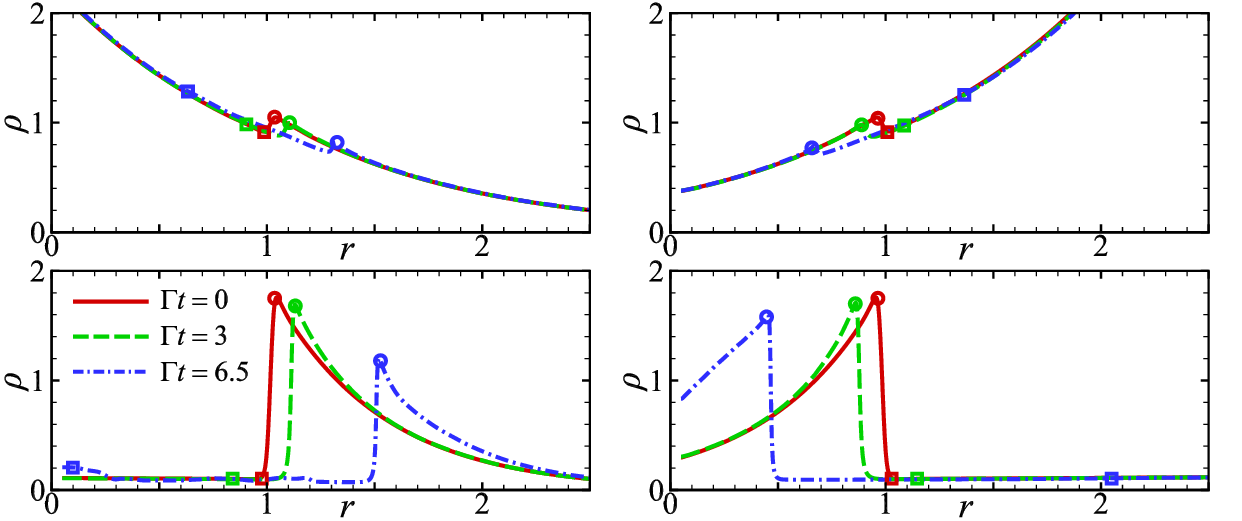}
			\put(-385,154){($e$)}
			\put(-195,154){($f$)}
			\put(-385,74){($g$)}
			\put(-195,74){($h$)}
			\caption{Density profiles along the radial lines across bubble tips at different $A_T$ with $n=8$ for ($a-d$) $Ma=0.1$ and for ($e-h$) $Ma=0.9$. The left column ($a,c,e,g$) and right column ($b,d,f,h$) denote convergent and divergent cases, respectively. The first and third rows ($a,b,e,f$) represent cases at $A_T=0.1$ and the second and last rows ($c,d,g,h$) denote cases at $A_T=0.9$. The circle/square marked on the density profile represents radial position of the bubble/spike tip at each moment ($\Gamma t=0,3,6.5$).}
			\label{fig:RLineRho}
		\end{figure}

		At $Ma=0.9$, the bubble accelerates more obviously at higher $A_T$ in highly nonlinear stage and varying $A_T$ leads to different bubble development trends compared with those at $Ma=0.1$. Specifically, decreasing $A_T$ from $0.9$ to $0.1$ suppresses the bubble evolution and this interesting phenomenon also appears in planar stratified RTI~\citep{Luo2020,Fu2022,Fu2023}. There are two mechanisms of the stabilizing effect of density stratification and the destabilizing effect of flow compressibility. They compete with each other, giving rise to different manifestations at different $A_T$. Figure~\ref{fig:RLineRho} shows density profiles along the radial lines across bubble tips for different parameters settings. At $A_T=0.1$, the density difference at the bubble tip suffers a decrease versus time for both convergent and divergent cases as shown in figure~\ref{fig:RLineRho}($e,f$). As a result, the stabilizing effect of density stratification becomes dominant at low $A_T$, leading to that $V_b$ starts to decay before reaching the saturated value. However, the density difference at the bubble tip reduces more slowly with the development of RTI at higher $A_T$. On the one hand, the intial density difference between two sides of the interface increases with the increase of $A_T$, thus weakening the initial density stratification. On the other hand, it is found from figure~\ref{fig:RLineRho}($g,h$) that the decreasing trend of density difference is greatly suppressed since the density before the bubble front has a significant increase compared to its initial state under the compression of heavy fluid exerted by the rising bubble~\citep{Luo2020,Fu2022}. Consequently, the stabilizing effect of density stratification gradually weakens while the destabilizing effect of flow compressibility strengthens with the increasing $A_T$, resulting in that the effect of $A_T$ on bubble behaviors at $Ma=0.9$ is quite different from those at $Ma=0.1$. Moreover, it is observed that there are profound differences between convergent and divergent cases for the bubble acceleration behaviors at $Ma=0.9$ where flow compressibility dominates the bubble acceleration behaviors. Specifically, $V_b$ for convergent cases increases robustly while keeps the fluctuation growth for divergent cases, indicating that the acceleration mechanism of flow compressibility is also fundamentally different in convergent and divergent cases.

		\begin{figure}
			\centering
			\includegraphics[scale=0.64]{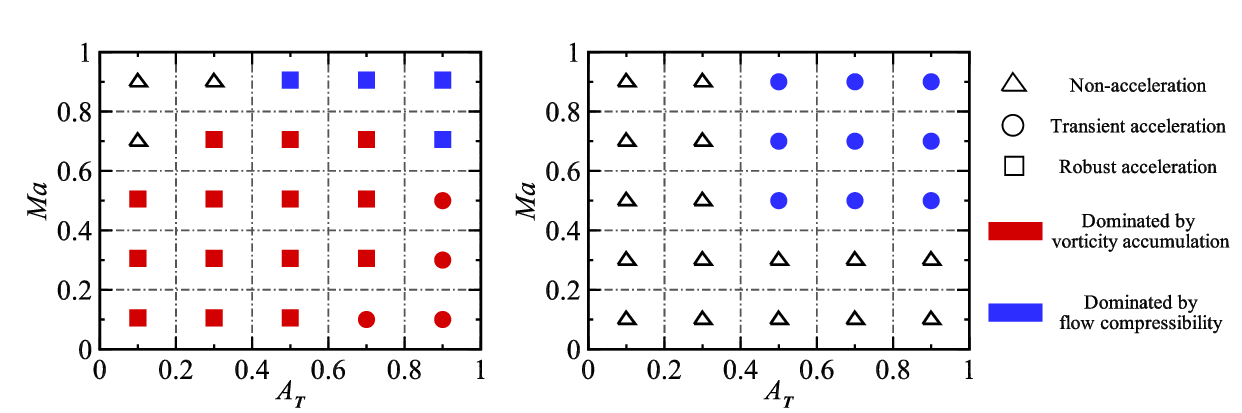}
			\put(-375,106){($a$)}
			\put(-225,106){($b$)}
			\caption{Bubble acceleration phase diagram of $A_T-Ma$ of ($a$) convergent and of ($b$) divergent cases for $n=8$ in highly nonlinear stage. Triangle, circle and square represent the phases of non-acceleration, transient acceleration and robust acceleration, respectively. Red and blue symbols indicate that the bubble acceleration mechanism is dominated by vorticity accumulation and flow compressibility, respectively.}
			\label{fig:Phase}
		\end{figure}

		Analogous to previous studies~\citep{Bian2020,Fu2023}, the highly nonlinear bubble behaviors are divided into three phases in figure~\ref{fig:Phase}: robust acceleration, transient acceleration and non-acceleration. In a robust acceleration phase, a bubble accelerates robustly at late time (e.g. convergent case with $A_T=0.1$ and $Ma=0.1$). In a transient acceleration phase, a bubble accelerates transiently but then decelerates (e.g. divergent case with $A_T=0.9$ and $Ma=0.9$). A non-acceleration phase denotes that bubble velocity starts to decay after reaching the asymptotic value~\citep{Goncharov2002} (e.g. divergent cases with $Ma=0.1$). Summarized in figure~\ref{fig:Phase}, a robust acceleration occurs in most $A_T-Ma$ space in convergent cases except for a non-acceleration of low $A_T$ and high $Ma$ cases, and a transient acceleration of high $A_T$ and low $Ma$ cases. Differently, a non-acceleration occurs for low $A_T$ or low $Ma$ and a transient acceleration appears at high $A_T$ and high $Ma$ without any robust acceleration phase in divergent cases. These results also vary considerably from the highly nonlinear bubble behaviors of planar stratified RTI~\citep{Fu2023}, which indicates that the traditional studies of late-time RTI performed in planar geometry cannot be robustly applied to the cylindrical or spherical counterparts.

	\subsection{Bubble growth model}
	    Based on potential flow assumption, \citet{Zhao2020b} derived a nonlinear ordinary differential equation (ODE) covering bubble evolution of incompressible cylindrical RTI with uniform-density background at arbitrary $A_T$ from linear to nonlinear regimes. The bubble amplitude $\eta(t)=r_b-r_0$ in convergent case satisfies the following evolution equation
		\begin{eqnarray}
			\begin{gathered}
			-\frac{n^2-4A_T H n-12A_T H^2}{2(6H-n)(r_0+\eta)}\frac{\mathrm{d}[(r_0+\eta)\dot\eta]}{\mathrm{d}t}-\frac{A_T H \dot\eta^2}{r_0+\eta}\\
			+\frac{(4A_T-3)n^2+6(3A_T-5)H n+36A_T H^2+12(A_T-1)H}{2(6H-n)^2 (r_0+\eta)}n^2\dot\eta^2=A_T g H.
			\end{gathered}
			\label{eq:ZhaoModel1}
		\end{eqnarray}
		Here, the parameter $H$ takes form as
		\begin{eqnarray}
			H=\left[\frac{n^2}{6n+4}-\frac{n^2\eta_0}{2(r_0+\eta_0)}\right]\left(\frac{r_0+\eta_0}{r_0+\eta}\right)^{3n+2}-\frac{n^2}{6n+4}.
			\label{eq:ZhaoModel1H}
		\end{eqnarray}
		Similarly, the evolution equation for the bubble amplitude $\eta(t)$ in divergent case is
		\begin{eqnarray}
			\begin{gathered}
			-\frac{n^2+4A_T H n-12A_T H^2}{2(6H+n)(r_0+\eta)}\frac{\mathrm{d}[(r_0+\eta)\dot\eta]}{\mathrm{d}t}-\frac{A_T H \dot\eta^2}{r_0+\eta}\\
			+\frac{(4A_T-3)n^2-6(3A_T-5)H n+36A_T H^2+12(A_T-1)H}{2(6H+n)^2 (r_0+\eta)}n^2\dot\eta^2=A_T g H.
		    \end{gathered}
			\label{eq:ZhaoModel2}
		\end{eqnarray}
		Here, the parameter $H$ is expressed as
		\begin{eqnarray}
			H=\left[-\frac{n^2}{6n-4}-\frac{n^2\eta_0}{2(r_0+\eta_0)}\right]\left(\frac{r_0+\eta}{r_0+\eta_0}\right)^{3n-2}+\frac{n^2}{6n-4}.
			\label{eq:ZhaoModel2H}
		\end{eqnarray}		
		However, the effects of density variation, vorticity accumulation and flow compressibility are not taken into consideration and thus this nonlinear theory~(\ref{eq:ZhaoModel1})--(\ref{eq:ZhaoModel2H}) cannot reasonably capture the nonlinear bubble behaviors of compressible cylindrical RTI. For example, this nonlinear theory~(\ref{eq:ZhaoModel1})--(\ref{eq:ZhaoModel2H}) correctly predicts the bubble behavior in the early nonlinear stage but fails in highly nonlinear phase when the flow compressibility dominates the flow field near the bubble at $A_T=0.9$ and $Ma=0.9$ as shown in figure~\ref{fig:Grid}($b$). Here, we attempt to improve this nonlinear theory~\citep{Zhao2020b} to quantitatively characterise the bubble growths of cylindrical RTI with isothermal stratification by considering effects of density variation, vorticity accumulation and flow compressibility.

		In groundbreaking work, \citet{Betti2006} took the effects of vorticity into account in planar ablative RTI and creatively modified the saturated velocity model~\citep{Goncharov2002} by introducing a vorticity term. \citet{Bian2020} further improved the highly nonlinear bubble growth model~\citep{Betti2006} by adding an efficiency factor $\eta=0.45$ to the vorticity term to account for the attenuation of vortices as they travel through the bubble tip region. The accumulated vorticity inside the bubble can additionally induce an equivalent bubble velocity of $\eta d_r \overline\omega/(2k)$ in the opposite direction of the external acceleration ($g$), where $\overline\omega=\int_V \rho |\omega| \mathrm{d} V/\int_V \rho\mathrm{d} V$ is the average vorticity in the volume ($V$) inside the bubble between the bubble vertex and the distance $1/k$ from the vertex into the bubble, and $d_r=\rho'_l/\rho'_h$ is the density ratio at bubble tip with $\rho'_h$ being the maximum density at the bubble vertex and $\rho'_l$ being the average density in the volume ($V$). Substituting temporally varied wavelength $k=2\pi/\lambda=n/r_b$ into $\eta d_r \overline\omega/(2k)$ can yield the equivalent velocity $V_{v}$ induced by the vorticity accumulation in cylindrical RTI as
		\begin{equation}
			V_{v}=\eta d_r \frac{\overline\omega}{2n/r_b}.
			\label{eq:Vorticity}
		\end{equation}
	    The efficiency factor $\eta=0.4$ is used here to account for the attenuation of vortices in cylindrical geometry. Notably, $d_r$ can illustrate the density variation at bubble tip and thus a local Atwood number $A'_T=(1-d_r)/(1+d_r)$ is introduced to the nonlinear ODE~(\ref{eq:ZhaoModel1})--(\ref{eq:ZhaoModel2H}). To this end, the equivalent velocity $V_{p}$ caused by the potential flow can be determined by means of a numerical solution of the nonlinear ODE~(\ref{eq:ZhaoModel1})--(\ref{eq:ZhaoModel2H}), i.e. $V_{p}=\dot\eta(t)$.

		By employing Green's formula, \citet{Fu2023} recently established a relation between dilatation ($\theta=\bnabla\bcdot\boldsymbol{u}$) and the velocity $V_{c}$ contributed by flow compressibility in planar compressible RTI. The similar strategy is employed here and integrating $\theta$ yields
		\begin{equation}
			\iint_S \theta \mathrm{d} S= \oint_{\p S} u^d_{\phi}\mathrm{d} r-u^d_r r\mathrm{d} \phi.
			\label{eq:Com1}
		\end{equation}		
		Here, $(u^d_{\phi},u^d_r)$ is the dilatational (irrotational) component of fluid velocity and $S$ is the annulus region where the heavy fluid is compressed by the bubble, from the bubble tip $r_b$ to its front position $r_{\infty}$ where the fluid keeps the hydrostatic equilibrium with the fluid velocity being zero. Naturally, interior or exterior boundary of the computational domain with the fluid velocity being almost zero is used as $r_{\infty}$ for divergent or convergent cases, respectively. The left-hand side of~(\ref{eq:Com1}) can be expressed by the spatial average dilatation $\overline\theta$ in the region $S$, i.e. $\pi (r^2_{\infty}-r^2_b)\overline\theta$. Given that a rising bubble of light fluid acts like a piston uniformly compressing its front heavy fluid~\citep{Luo2020,Fu2022,Fu2023}, the compressing velocity $u^d_r$ at the radial position $r_b$ is almost uniform along the periodic $\phi$-direction and can be approximated as $V_{c}$~\citep{Fu2023}. Thus, the right-hand side of~(\ref{eq:Com1}) can be approximately calculated as $-2\pi r_b V_{c}$. Therefore, the equivalent velocity $V_{c}$ contributed by flow compressibility is modelled as
		\begin{equation}
			V_{c}=-\overline\theta(r_{\infty}-r_b)\frac{1+r_{\infty}/r_b}{2}.
			\label{eq:Com2}
		\end{equation}			
		Particularly, the factor $(1+r_{\infty}/r_b)/2$ in~(\ref{eq:Com2}) characterises the effect of interfacial curvature on flow compressibility. When the interfacial curvature doesn't matter, i.e. $r_b\rightarrow r_{\infty}$, the factor $(1+r_{\infty}/r_b)/2\rightarrow 1$ makes~(\ref{eq:Com2}) recover the velocity contributed by flow compressibility in planar compressible RTI~\citep{Fu2023}.

\begin{figure}
			\centering
			\includegraphics[scale=0.64]{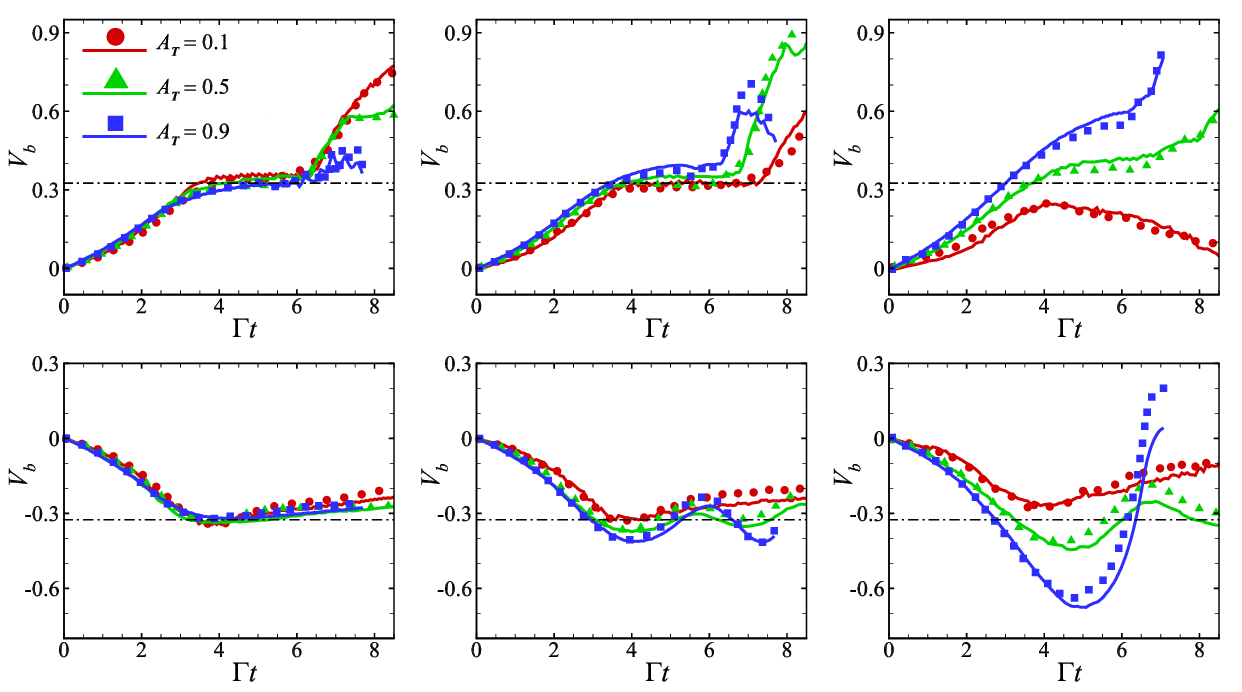}
			\put(-385,203){($a$)}
			\put(-258,203){($b$)}
			\put(-131,203){($c$)}
			\put(-385,97){($d$)}
			\put(-258,97){($e$)}
			\put(-131,97){($f$)}
			\caption{Bubble velocities obtained from DNS results (symbols) and calculated by the present model (lines) for different $A_T$ at ($a,d$) $Ma=0.1$, ($b,e$) $Ma=0.5$ and ($c,f$) $Ma=0.9$. The top row ($a-c$) and bottom row ($d-f$) denote convergent and divergent cases for $n=8$, respectively. The horizontal black dot-dashed lines in each panel denote the saturation velocity obtained from potential flow theory~\citep{Goncharov2002}.}
			\label{fig:Model1}
		\end{figure}
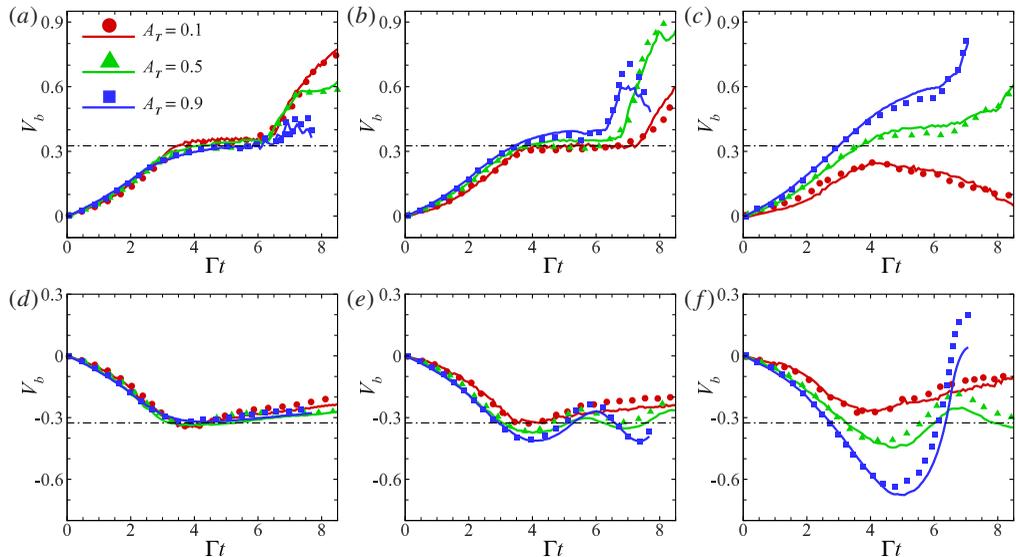

		The $V_{p}$ based on the potential flow assumption and density variation at bubble tip excludes the effects of vorticity accumulation and flow compressibility. By adding (\ref{eq:Vorticity}) and (\ref{eq:Com2}) to $V_{p}$, the bubble velocity can be completed as the following model:
		\begin{equation}
			V_{b}=V_{p}+V_{v}+V_{c}.
			\label{eq:Model}
		\end{equation}
		To verify the above improved model, we employ DNS on compressible cylindrical RTI over a wide range of Atwood numbers ($A_T=0.1-0.9$) and Mach numbers ($Ma=0.1-0.9$), especially in the nonlinear and highly nonlinear regimes. Figure~\ref{fig:Model1} shows bubble velocities obtained from DNS results (symbols) and calculated by this improved model (lines) for convergent (top row) and divergent (bottom row) cases. It is identified that this improved model~(\ref{eq:Model}) well reproduces the bubble velocity development from linear to highly nonlinear regimes.

		To further elucidate the acceleration mechanism in the highly nonlinear regime, the values of $\langle V_{v}- V_{v}|_{t_s}\rangle$ and $\langle V_{c}- V_{c}|_{t_s}\rangle$ roughly correspond to the contributions of vorticity accumulation and flow compressibility, respectively, following closely previous work~\citep{Fu2023}. Here, $\langle\bcdot\rangle$ denotes the time average from the moment $t_s$ when the bubble velocity reaches the value predicted by \citet{Goncharov2002} to the end of simulation. $V_{v}|_{t_s}$ and $V_{c}|_{t_s}$ denote the values of $V_{v}$ and $V_{c}$ at time $t_s$, respectively. It is clearly seen in figure~\ref{fig:Phase} that the acceleration in convergent cases is dominated by vorticity accumulation at low $A_T$ and low $Ma$ and by flow compressibility at high $A_T$ and high $Ma$ whereas the acceleration in divergent cases is purely induced by flow compressibility. Besides, the $A_T-Ma$ space of acceleration dominated by vorticity accumulation becomes greatly broader in convergent cases than that in corresponding planar counterparts with the same $Re_p$~\citep{Fu2023} and does not appear in divergent cases, and the $A_T-Ma$ space of acceleration induced by flow compressibility is much narrower in convergent cases than that in divergent cases.

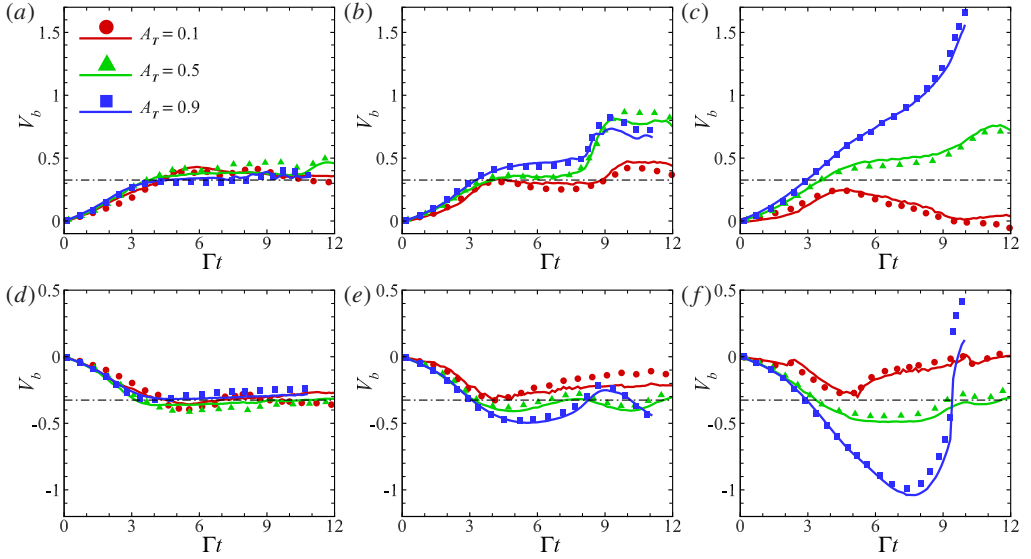
\begin{figure}
			\centering
			\includegraphics[scale=0.64]{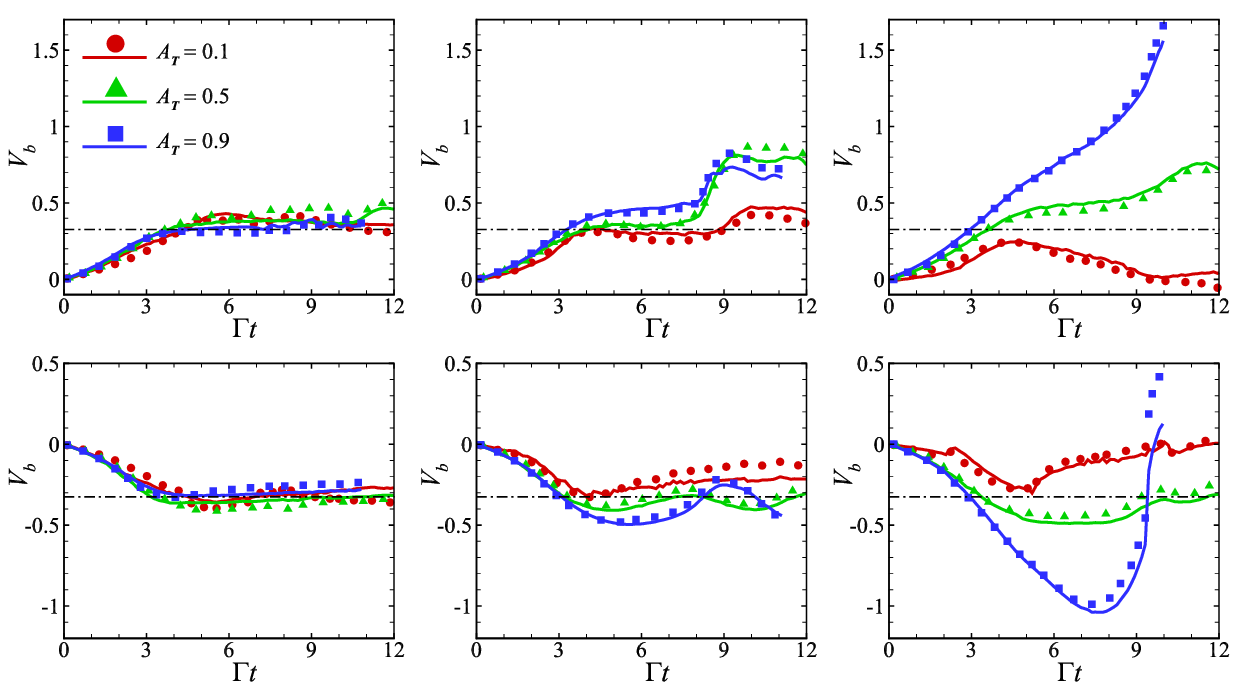}
			\put(-385,203){($a$)}
			\put(-258,203){($b$)}
			\put(-131,203){($c$)}
			\put(-385,97){($d$)}
			\put(-258,97){($e$)}
			\put(-131,97){($f$)}
			\caption{Bubble velocities obtained from DNS results (symbols) and calculated by the present model (lines) for different $A_T$ at ($a,d$) $Ma=0.1$, ($b,e$) $Ma=0.5$ and ($c,f$) $Ma=0.9$. The top row ($a-c$) and bottom row ($d-f$) denote convergent and divergent cases for $n=16$, respectively. The horizontal black dot-dashed lines in each panel denote the saturation velocity obtained from potential flow theory~\citep{Goncharov2002}.}
			\label{fig:Model2}
		\end{figure}

		To further validate the present improved model~(\ref{eq:Model}), both convergent and divergent cases for $n=16$ are simulated. Other parameters have kept the same as those of corresponding cases for $n=8$. It is clearly shown in figure~\ref{fig:Model2} that the present model can also well describe the compressible cylindrical RTI evolution of the cases for $n=16$, indicating the reliability of the present characterision of bubble evolution for single-mode cylindrical RTI with isothermal stratification.

\section{Concluding remarks}
    In this paper, nonlinear and highly nonlinear bubble evolutions of two-dimensional single-mode RTI with isothermal stratification are investigated in cylindrical geometry via DNS for different Atwood numbers ($A_T=0.1-0.9$) and Mach numbers ($Ma=0.1-0.9$). It is found that the nonlinear bubble growth involves the effects of density stratification, vorticity accumulation and flow compressibility and shows considerable differences between convergent and divergent cases. Specifically, strong stabilizing effect of density stratification leads to non-acceleration at low $A_T$ and high $Ma$. The accelerations in convergent cases are dominated by vorticity accumulation at low $A_T$ and low $Ma$ and by flow compressibility at high $A_T$ and high $Ma$ whereas the accelerations in divergent cases are purely induced by flow compressibility at high $A_T$ and high $Ma$. Moreover, the $A_T-Ma$ space of acceleration dominated by vorticity accumulation becomes broader in convergent cases than that in corresponding planar counterparts~\citep{Fu2023} and does not appear in divergent cases. And the acceleration dominated by flow compressibility is always robust in convergent cases while transient in divergent cases.

	Based on the nonlinear theory~\citep{Zhao2020b} of incompressible cylindrical RTI with uniform-density background, an improved model has been proposed to describe the bubble velocity development by taking density variation, vorticity accumulation and flow compressibility into consideration. This model well reproduces the bubble velocity development from linear to highly nonlinear regimes. It is noted that current model inheriting the illuminating idea of the Betti-Sanz model~\citep{Betti2006} requires the simulation and thus is not predictive. Nevertheless, it will be helpful to understand the physical mechanisms of compressible RTI in cylindrical geometry. The present work sheds light on the mechanisms of two-dimensional single-mode RTI. If insight is further taken into the behaviour of compressible multi-mode cylindrical RTI, it is possible to better understand the evolution of the resultant turbulent mixing.
	
\backsection[Acknowledgements]{The authors are very grateful to Dr. Y.-S. Zhang at the Institute of Applied Physics and Computational Mathematics for useful discussions on the algorithm and code.}

\backsection[Funding]{This work was supported by the National Natural Science Foundation of China (Nos. 12202436, 12388101, 12293000 and 12293002), by Science Challenge Project, by LCP Fund for Young Scholar (No. 6142A05QN22002) and by the Strategic Priority Research Program of the Chinese Academy of Sciences (No. XDB0500301)}

\backsection[Declaration of interests]{The authors report no conflict of interest.}

\backsection[Author ORCIDs]{
	
	Ming Yuan, https://orcid.org/0000-0002-2602-9990;
	
	Zhiye Zhao, https://orcid.org/0000-0003-1509-5084;
	
	Luoqin Liu, https://orcid.org/0000-0002-6020-3702;
	
	Nan-Sheng Liu, https://orcid.org/0000-0001-9128-1933;
	
	Xi-Yun Lu, https://orcid.org/0000-0002-0737-6460.}
		
\bibliographystyle{jfm}
\bibliography{jfm}
%Use of the above commands will create a bibliography using the .bib file.

\end{document}